\documentclass[preprint,aps,showpacs,superscriptaddress,preprintnumbers,amsmath,amssymb]{revtex4}

\usepackage{graphicx}
\usepackage{dcolumn}
\usepackage{bm}

\newcommand{\rms}{r_{\rm nn}^{\rm rms}}

\begin{document}

\title{Neutron-Neutron Correlations in the Dissociation of Halo Nuclei \footnote{Revised and updated  version of an invited contribution to the RCNP Osaka Spring Workshop
on Cluster Condensation and Nucleon Correlations in Nuclei}}

\author{N.A.~Orr}
\affiliation{LPC-Caen, ENSICAEN, Universit\'e de Caen et IN2P3-CNRS, 14050 Caen Cedex, France}

\begin{abstract}
Studies attempting to probe the spatial configuration of the valence neutrons in 
two-neutron halo nuclei using the technique of intensity 
interferometry are described.
Following a brief review of the method and its application to 
earlier measurements of the 
breakup of $^{6}$He, $^{11}$Li and $^{14}$Be, the results of the analysis
of a high statistics data set
for $^{6}$He are presented.  The limitations of the technique, including the
assumption of incoherent emission in the breakup and the sensitivity
to the continuum states populated in the dissociation rather than the ground state, are discussed.

\end{abstract}

\pacs{25.75.Gz; 21.10.Gv; 27.20.+n}

\maketitle

\section{Introduction}	

Clustering, which has long been known to occur along the line
of $\beta$-stability
also appears in exotic forms as the drip-lines are approached. 
The most spatially extreme form of clustering is that
exhibited by neutron haloes
which appear as the ground states of very weakly bound nuclei at the limits of particle 
stability. Perhaps the most intriguing 
of the halo systems are the two-neutron halo nuclei $^{6,8}$He, 
$^{11}$Li, 
$^{14}$Be and $^{17}$B, in which the two-body subsystems -- core-$n$ and $n$-$n$ -- are 
unbound. Such 
Borromean behaviour naturally gives rise to the question of the correlations between the 
constituents. In particular, a long standing issue has been the
spatial configuration of the two halo neutrons.  Indeed, from an historical perspective, the
first attempts to model $^{11}$Li employed a core + point dineutron description~\cite{Han87}.  Later, more 
sophisticated three-body modelling lead to the concept of both dineutron-like and ``cigar''-type 
configurations~\cite{Zhu93}, a theme which has resurfaced in recent years~\cite{Hagino}.

Experimentally the probing of the correlations between the halo neutrons has proved
difficult.  Attempts have been made to employ two-neutron transfer reactions where, 
quite apart from the availability of relatively intense beams and the uncertainties 
in the optial model potentials, the
principal difficulty is the need to accurately model 
both pair and sequential transfer.  Such an approach for $^{6}$He has
suggested that, within the constraints of the three-body wavefunction used, pair
transfer of a dineutron-like
configuration dominates~\cite{Oga}.

High-energy dissociation reactions have also been explored as a possible probe.  
Experimentally the very high cross sections and thick targets allow relatively weak beams to be employed
and hence the most neutron-rich systems to be studied.  Some attempts have
been made to relate the measured neutron-neutron momentum or relative-angle distributions to the
halo neutron correlations in $^{11}$Li~\cite{Iek96,Esb93}.  More recently, Nakamura {\em et al.}
have derived an estimate, based on sum-rule considerations, of the $n$-$n$ opening angle from a measurement of the dipole strength function in the dissociation of
$^{11}$Li~\cite{Nak06}.  Motivated by this work, Hagino and Sagawa have proposed a less model dependent analysis
based on the dipole strength function and the total matter radius~\cite{Hag07}.

In the work described here, we have applied the
technique of intensity interferometry and the corresponding correlation function to analyse the  neutron-neutron relative momentum ($q$)
distributions measured following dissociation at intermediate energies.  Assuming that incoherent (ie., independent particle) emission is a reasonable
approximation for the dissociation of halo systems, and using an analytical description 
of the $n$-$n$ correlation function with a
Gaussian source parameterisation, estimates of the RMS neutron-neutron 
separation ($r_{nn}^{rms}$) at breakup 
have been derived.
It is worthwhile noting at the outset that, within these limitations, the technique is 
sensitive to the configurations present in the states populated in the continuum by 
the dissocation rather than the ground state of the projectile.  As such, the use of the 
$r_{nn}^{rms}$ so obtained to deduce, in a ``model independent'' manner, the ``geometry'' of
two-neutron halo systems~\cite{Mahir,Hag07} should be proscribed.

It should be stressed that the correlation function analysis presented here is  
an approximation of sorts to the approach which should in principle be employed; namely a
comparison of the measured $n$-$n$ relative momenta to the results of 
a complete calculation based on a realistic
halo wavefunction coupled to a full reaction calculation incorporating the final-state 
interactions in the outgoing channel.  
To date, however, such calculations have,
with the exception of the pioneering work of Esbensen {\em et al.} on the Coulomb breakup of
$^{11}$Li~\cite{Esb93}, yet to be attempted.

\section{Technique}

\begin{figure}[th]
\includegraphics[width=14cm]{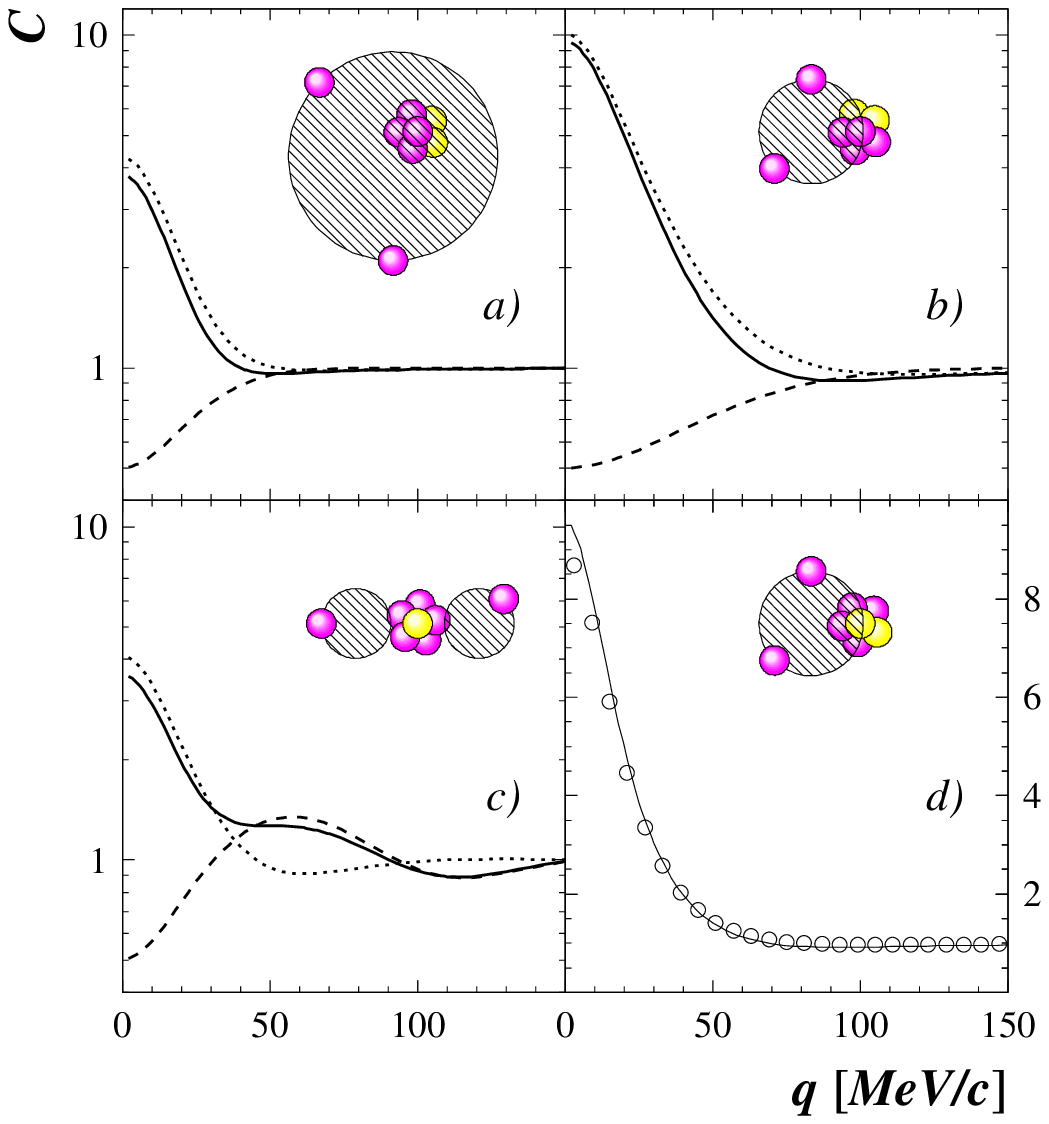}
\caption{Neutron-neutron correlation functions, $C$, for different halo 
configurations (note the logarithmic scale for $C$).  The calculations 
are based on Gaussian sources (see text) with sizes,
$r_0$, of (a) 6 fm, (b) 3 fm and (c) 2 fm separated by 10 fm.  The 
contributions from Fermi--Dirac statistics and the neutron--neutron FSI are 
shown by the dashed and dotted lines respectively. 
Simultaneous emission for a source size of
3 fm (solid line) is compared in (d) to a source with a space-time extent of 3~fm, 
50~fm/c
(open symbols)~\protect\cite{FMM00}.}  
\label{f:fig1}  
\end{figure}

Intensity interferometry was developed as part of the pioneering work on stellar 
interferometry (using optical wavelength photons) of
Hanbury-Brown and Twiss in 
the 1950's and 60's~\cite{HBT}.  The method was later extended to source size measurements using pions 
in high-energy physics~\cite{Gol60}.  The principle behind the technique is as follows:  
when identical particles are emitted 
in close proximity in space-time, the wave function
of relative motion is modified by the FSI and quantum statistical
symmetries (QSS)~\cite{Boa90} --- in the case of halo neutrons the overwhelming effect is 
expected to be that of 
the FSI~\cite{FMM00} (Fig.~\ref{f:fig1}). Intensity interferometry relates this
modification to the space-time separation of the particles at emission as a 
function of the four-momenta of the particles through the correlation function 
$C_{\rm{nn}}$, which is defined as,

\begin{equation}
 C_{\rm{nn}}(p_1,p_2)=\frac{d^2n/dp_1dp_2}{(dn/dp_1)\,(dn/dp_2)} \label{e:C12}
\end{equation}

\vspace*{10pt}

where the numerator is the measured two-particle distribution and the denominator the 
product of the independent single-particle distributions~\cite{FMM00} -- that is, the distribution
a particle would exhibit if it were not influenced by the other.  
As is generally the case, the single-particle distributions
have been generated via event mixing.  Importantly, in the 
case of halo neutrons
special consideration must be given to the strong ``residual correlations'' 
which are present in the event-mixed distributions~\cite{FMM00}.  Here, an iterative technique developed in the
spirit of that employed by Zajc {\em et al.}~\cite{Zajc} was used to produce the final event-mixed distributon~\cite{FMM00}. The neglect of
the effects engendered by the residual correlations can result in a significant overestimate of the corresponding apparent 
source size~\footnote{Interestingly,
the first attempt to construct $C_{\rm{nn}}$ for the breakup of
a halo nucleus ($^{11}$Li) not only failed to recognise the problem of
residual correlations, but also interpreted the source size as a direct measure of the 
overall size of the halo~\cite{Iek93}.}.
Experimentally much care needs to be taken to eliminate
cross talk (ie., a single scattered neutron that is registered in two
detectors), which will tend to enhance the number of events at low-relative momentum
and hence artifically diminish the apparent source size~\cite{Mar00}.

In the case of neutron pairs, the QSS are governed by Fermi statistics and the
FSI by the strong interaction. In our work we have employed 
the formalism developed by Lednicky and Lyuboshits~\cite{soviet},
where analytic expressions for the neutron-neutron correlation function 
in the s-wave channel, with 
explicit dependence on
the distribution of their relative four-distance, were derived. 
In Fig.~\ref{f:fig1} the sensitivity of $C$ to different
halo-neutron configurations is demonstrated, where for simplicity Gaussian
source parameterisations have been used. As pointed out in Ref.~\cite{soviet}, the repulsive 
contribution of the QSS is overwhelmed at low
$q$ by the attractive FSI, and $C$ may take on much higher values --
$C(0)\sim10$--20 -- than is possible in other cases (eg., bosons or
charged fermions~\cite{Boa90,PRP} or evaporated neutrons with emission over a long
time scale~\cite{compound}), where $C(0)<2$. Two-neutron interferometry thus
presents the additional advantage for probing haloes that it may, within 
the context of the approximations outlined here, be applied to
data sets of relatively poor statistical quality -- an important feature given
the generally low beam intensities and triple coincidence requirement 
of the experiments.

\section{Results and Discussion}

\begin{figure}[th]
\includegraphics[width=13cm]{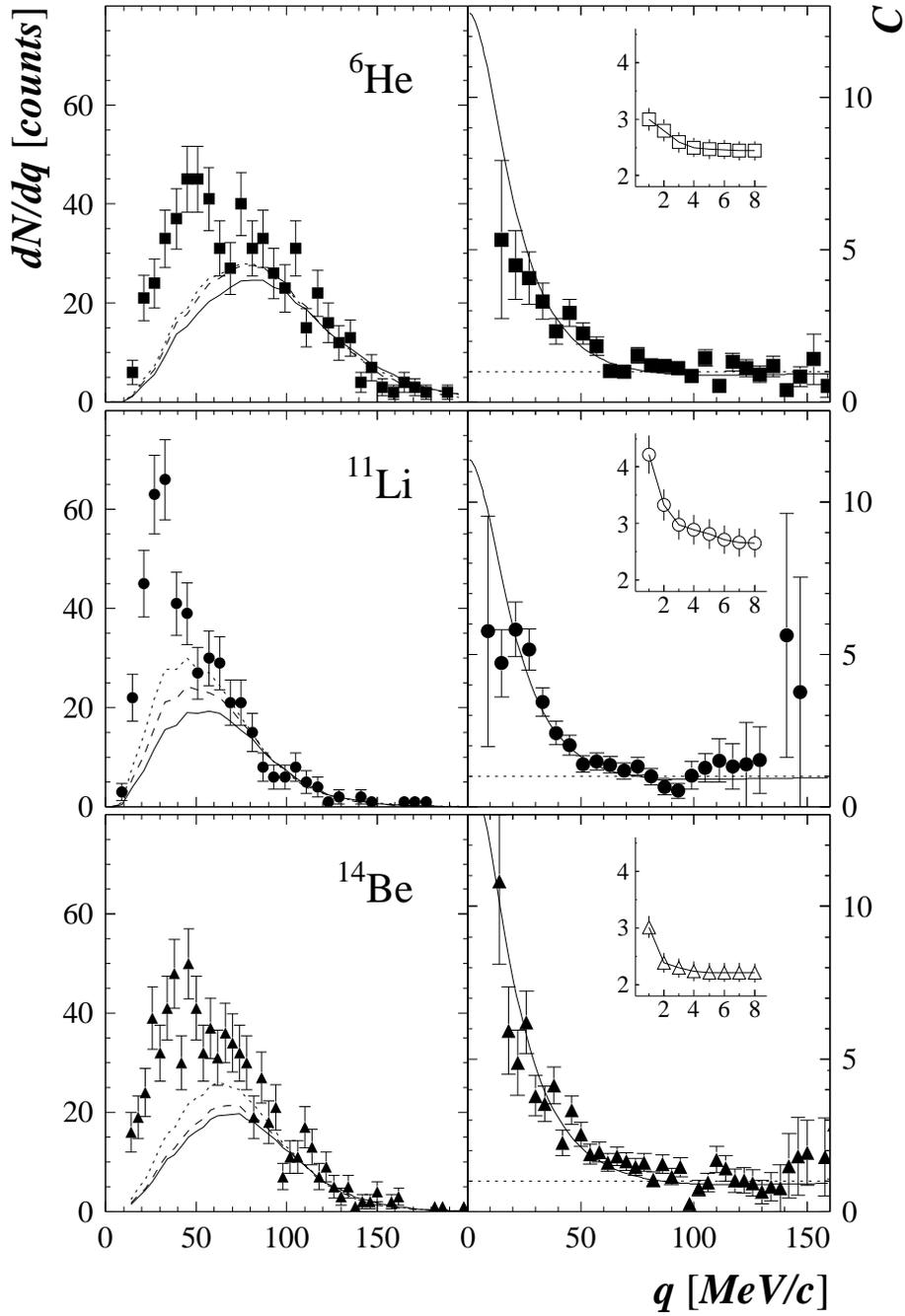} 
 \caption{Correlation functions (right hand panels) constructed for the 
dissociation of the halo nuclei $^{6}$He, $^{11}$Li and 
$^{14}$Be by a Pb target. The solid lines correspond to the fit 
based on a Gaussian source (see text). On the left, the measured numerator (symbols) and 
the successively reconstructed event mixed distributions (dotted, dashed and solid lines for 
$i=1,2,8$, respectively) are displayed -- see Ref.~\protect\cite{FMM00} for more details. The insets 
demonstrate the evolution of the source size ($r_0$ fm) with the number of iterations.}
 \label{f:fig2}
\end{figure}

As a first step, measurements of the dissociation by a Pb target
of intermediate energy beams of $^{6}$He, $^{11}$Li and 
$^{14}$Be to the core-$n$-$n$~\cite{Lab01} were analysed~\cite{FMM00} (the details of 
the experiment may be found in
Refs~\cite{FMM00,Lab01}).  The choice of data acquired with a high-Z target was made to 
privilege
Coulomb induced breakup, whereby, in a scenario similar to that argued for by by Ieki {\em et al.}~\cite{Iek93}, the halo neutrons were, to a first approximation,
expected to act as {\em spectators}, and for which simultaneous emission might be 
assumed to occur (see below).  The correlation functions derived from the data (Fig.~\ref{f:fig2}), 
assuming simultaneous emission, were compared to
the analytical formalism based on a Gaussian source~\footnote{For a Gaussian source of
variance $r_0$ (Eq.~(24) Ref.~\cite{soviet}) the n-n separation $r_{\rm nn}$ is 
a Gaussian distribution of variance 
$\sqrt{2}\,r_0$ and $\rms=\sqrt{6}\,r_0$.}.
Such a parametrization describes well the experimental correlation functions and
neutron-neutron separations at breakup of $r_{nn}^{rms}=5.9\pm1.2$~fm ($^{6}$He), 
$6.6\pm1.5$~fm ($^{11}$Li) and $5.6\pm1.0$~fm ($^{14}$Be) were thus deduced.  These 
results appeared, within the limitations of the technique, to
preclude any strong compact dineutron component at breakup; a
result which for $^{6}$He, assuming that the ``spectator'' hypothesis is valid, 
would be in line with a complementary radiative capture study reported 
by us in Ref.~\cite{Sau00}.
It is interesting to compare these results to the RMS neutron-proton
separation of 3.8~fm in the deuteron (the only bound two-nucleon system).

The same analysis was applied to dissociation of $^{14}$Be by a C target,
in order to investigate the influence of the reaction mechanism~\cite{FMM01}.  A 
result which hinted at a somewhat larger separation at breakup, $r_{nn}^{rms}=7.6\pm1.7$~fm, was 
obtained. This raised the issue as to whether simultaneous emission could be
assumed a priori~\cite{FMM01}. The three-body nature of the system breaking up suggested
that any delay in the emission of one of the neutrons would arise
from core-$n$ FSI/resonances in the exit channel; a process that might 
be expected to be
enhanced in nuclear induced breakup owing to the significant
contribution to dissociation attributable to diffractive breakup~\cite{Lab01}. 

From a pragmatic point of view, given the statistical 
quality of the data, the relatively simple (Gaussian) parameterisation of the 
neutron source and the other approximations inherent in the technique, 
such considerations are probably not yet relevent.  Moreover, as demonstrated by the
comparison presented in Fig.~\ref{f:fig1}d, the difference between correlation
functions (for a source with $r_0$=3~fm) for simultaneous 
emission and passage via an intermediate resonance with a lifetime of 
50~fm/c ($\Gamma \sim$~4~MeV) is negligible.  Indeed, a lifetime of order 1000~fm/c or less
($\Gamma \leq $~0.2~MeV) would be required before the effects would be significant for
the interpretation of the data presented here.

In addition to the influence of such two-step breakup, it must also be recognised,
as pointed out earlier, that
the neutron-neutron configuration being probed is not that of the projectile
ground state but rather the average over the states in the continuum populated in the
breakup which subsequently decay to the core-$n$-$n$ channel.  Indeed, it was in this spirit 
that our first analyses (as described above) were undertaken of data acquired for  
Coulomb dominated breakup, whereby dipole excitations with a large overlap
with the ground state $n$-$n$ configuration were argued to occur.

In order to explore in more detail these issues and the technique in general, 
a dedicated study of the breakup of $^{6}$He was undertaken, the details of which may be 
found in Ref.~\cite{Guillaume}.  Experimentally $^{6}$He presents the advantage
that it can be produced as a relatively intense beam 
(here $\sim$3$\times$10$^{4}$~pps~\footnote{In the 
present case the intensity was limited by the count rate capabilities of 
the segmented zero-degree telescope 
used to detect the charged fragments and beam~\cite{StevePRL}.} at
30~MeV/nucleon).  Theoretically $^{6}$He is the two-neutron halo nucleus which 
can be modelled
the most reliably and for which the core-$n$ system, $^{5}$He~\footnote{The ground 
state being a resonance some 0.8~MeV above 
threshold with a width
$\Gamma \approx$ 0.7~MeV -- the latter corresponding to a lifetime short enough to avoid
any need to consider the implications of sequential decay.}, is the most well established .  

\begin{figure}[th]
\includegraphics[width=14cm]{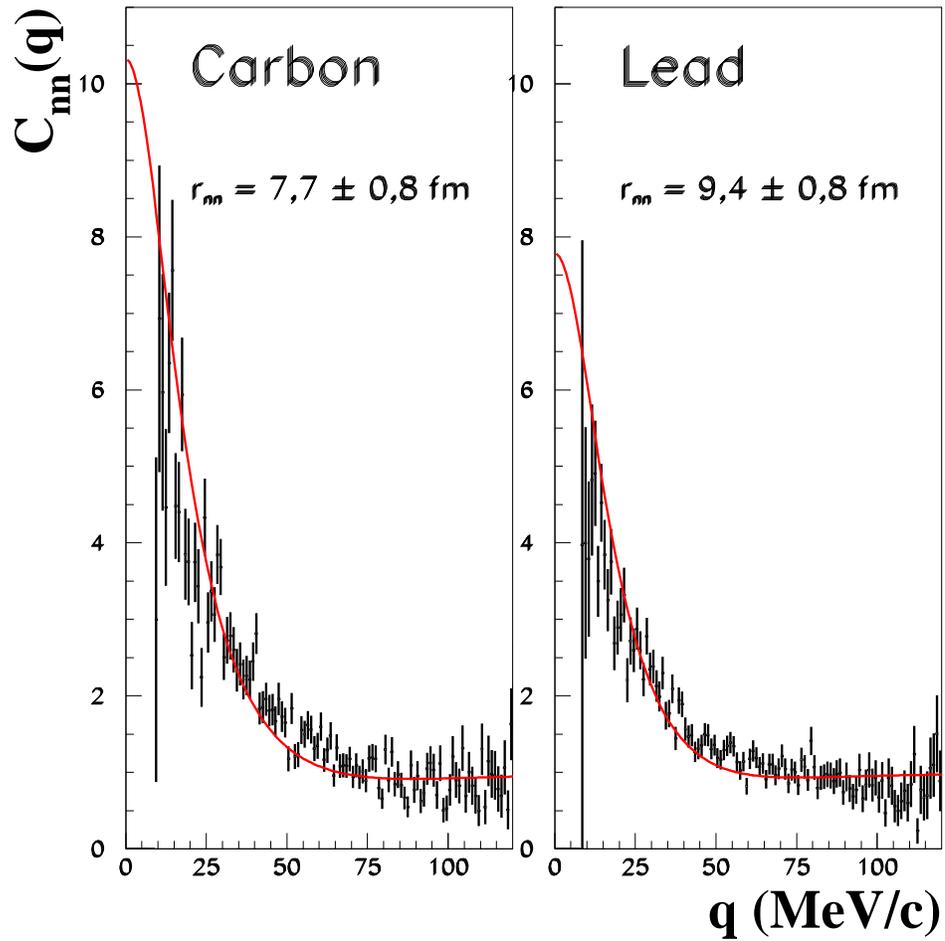}
 \caption{Correlation functions constructed for the 
dissociation of $^{6}$He by C~\protect and Pb targets~\protect\cite{Guillaume}. The solid lines 
correspond to the fit 
assuming a Gaussian source (see text) and the corresponding $r_{\rm nn}^{\rm rms}$ are indicated.}
\label{f:fig3}
\end{figure}

The correlation functions derived from the data acquired for reactions on both C 
and Pb targets are displayed in Fig.~\ref{f:fig3}.  As in our earlier work, a Gaussian source
parameterisation based on the formalism of Ref.~\cite{soviet}, was employed to 
derive estimates of the RMS $n$-$n$ separation at breakup of
7.7$\pm$0.8 (C) and 9.4$\pm$0.8~fm (Pb) for the reactions on the two targets.  As noted 
above, the $n$-$n$ configurations being probed are 
those of the unbound states in $^{6}$He populated in the reaction which subsequently decay
to $\alpha$-$n$-$n$.  The difference in the apparent source sizes presumably arises 
then from a
different selectivity in the states populated in the two cases.  To pursue this
conjecture further the decay energy spectrum (E$_d$ = E$_x$ - S$_{2n}$) 
for the $^{6}$He$^*$ have been reconstructed 
from the measured
momenta of the $\alpha$ and two neutrons.  

\begin{figure}[th]
\includegraphics[width=14cm]{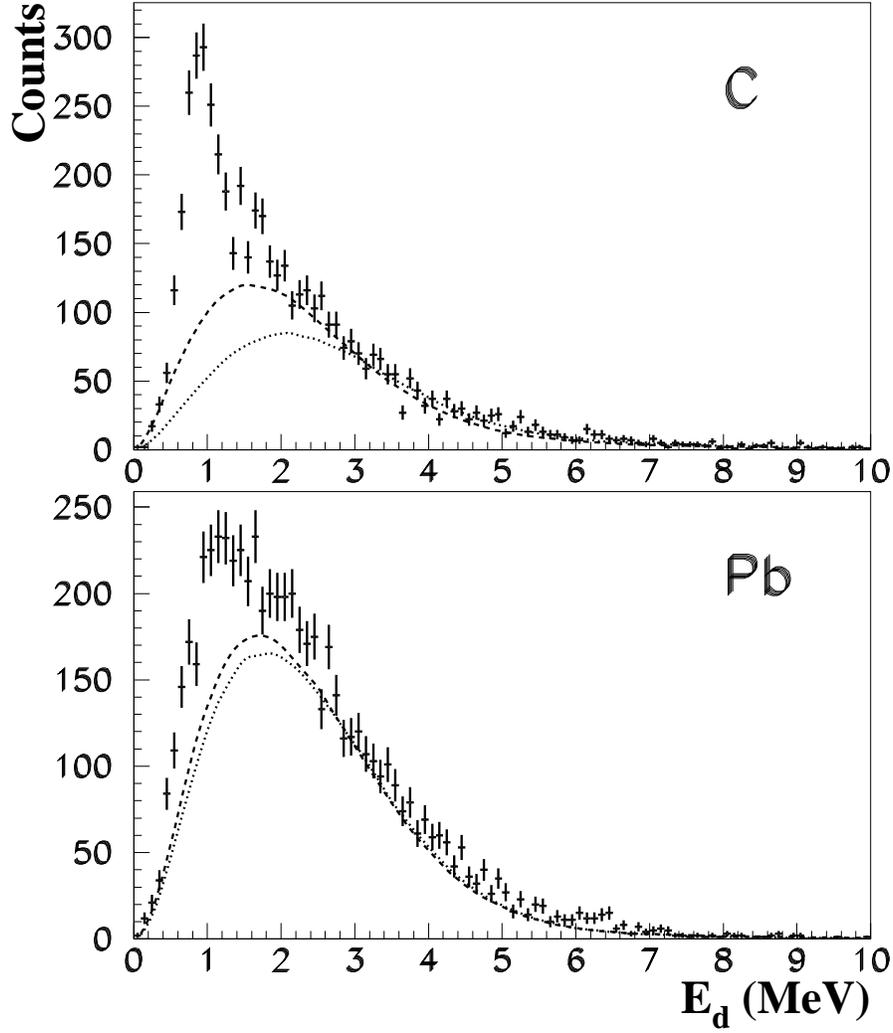}
 \caption{Decay-energy spectra for the 
dissociation of $^{6}$He by C and Pb targets~\protect\cite{Guillaume}. The 
relatively narrow peak at E$_d \sim$1~MeV is the
well known 2$^+_1$ state (E$_x$=1.8~MeV).  Uncorrelated $\alpha-n-n$ distributions 
generated by event
mixing and normalised to the data at high E$_d$ are shown as the dashed 
and dotted lines (first and final iterations 
respectively -- see Ref.~\protect\cite{Guillaume}).}
\label{f:fig4}
\end{figure}

As may be seen in Fig.~\ref{f:fig4}, the spectra are markedly different -- that arising from reactions 
on the C target is dominated by the well established 2$^+_1$ state at E$_x$=1.8~MeV, whilst 
in the case of the Pb target the spectrum is relatively featureless, with the 2$^+_1$ state 
being much more weakly populated~\footnote{Broadly similar results have been found 
by the LAND Collaboration in measurements made at a beam energy an order of 
magnitude higher than that employed here~\cite{Aum99}.}.  Given that breakup on the Pb 
target would be expected to be 
dominated by Coulomb induced reactions, the excitation-energy spectrum should 
contain significant E1 strength.  Interestingly, three-body 
models predict an E1 distribution
with a peak at around 1-1.5~MeV decay energy (a feature hinted at in Figs 4 and 5) and 
a strong, slowly decaying ``tail'' towards
higher energies~\cite{Cob97,Dan98}.

\begin{figure}[th]
\includegraphics[width=15.5cm]{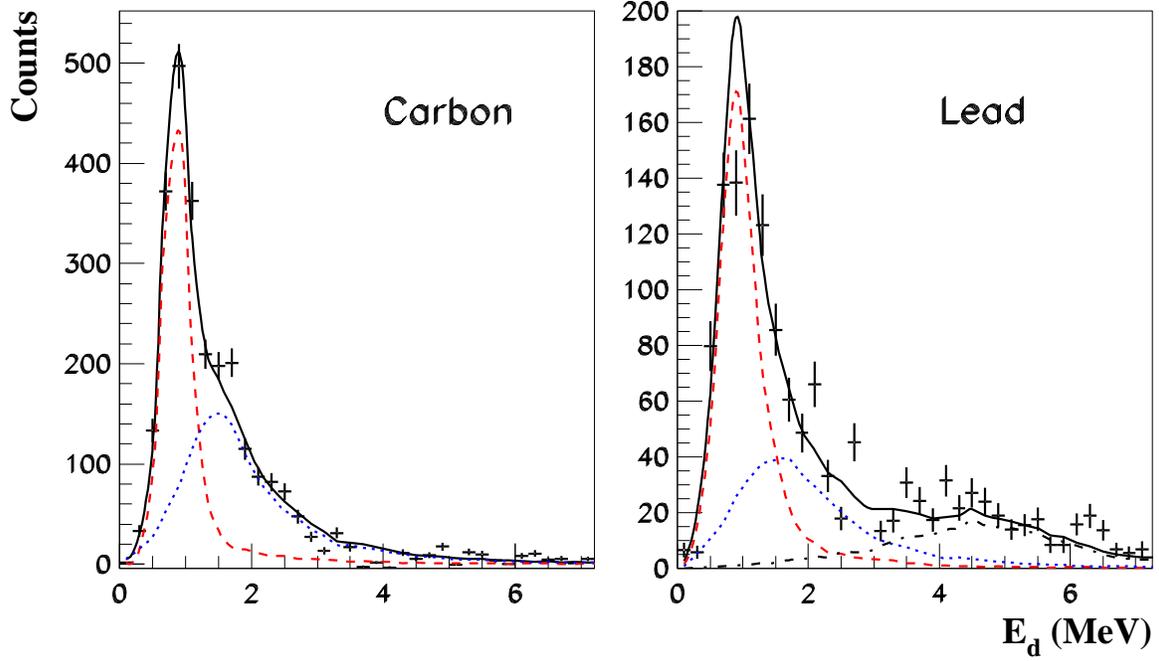} 
 \caption{Decay-energy spectra for the 
dissociation of $^{6}$He by C and Pb targets after subtraction of the uncorrelated $\alpha-n-n$ distribution
obtained via event mixing (dotted distributions in Fig.~\protect\ref{f:fig4}).  The resonance parameters 
used for the three components shown
are~\protect\cite{Guillaume}: E$_r$ = 0.82, 1.5, 4~MeV; $\Gamma$ = 0.16, 0.5, 6~MeV.}
\label{f:fig5}
\end{figure}

In Fig.~\ref{f:fig5} the decay energy spectra are shown after subtraction of the uncorrelated
distribution generated by event mixing.  As the E1 strength is not expected to be associated with 
well defined three-body resonances~\cite{Dan04}, such a comparison should suppress it and any
other non-resonant structures.  In this context, we note that the calculations of 
Danilin {\em et al.}~\cite{Dan04}
suggest that the monopole strength, which may be expected to be preferentially populated
by reactions on the C target, does not present a strongly correlated character either.

Danilin {\em et al.}~\cite{Dan04} have investigated theoretically 
spatial correlations in the three-body continuum
of $^{6}$He and conclude that the 2$^+_1$ state presents a relatively compact structure --  
with a ``most probable'' n-n separation of $\sim$5~fm -- compared to the $1^-$ continuum states.
This is qualitatively in line with the results obtained here of a larger $r_{\rm nn}$ at breakup for 
reactions on the Pb target
as compared to the C target.

\section{Conclusions}

To proceed in a more quantitative manner, $n$-$n$ correlation functions derived from
realistic halo wavefunctions, weighted by the
relative yields expected in the reactions under study, need to be constructed.  Furthermore, the
degree to which coherent emission affects the results should be explored.   The first steps in this direction have recently been made 
by Yamashita {\em et al.}~\cite{Yam05}, whereby these issues have been addressed for the first time.
Interestingly, the shapes of the correlation functions at low relative momenta 
($q$~$\lesssim$~80~MeV/c) are in reasonably good agreement
with the data and with the simple formalism based on incoherent emission and Gaussian source parameterisations
employed here.  It may be speculated that the differences observed at large 
relative momenta arise from the $l > $0 components of the $n$-$n$ interaction which were not included in the formalism usec to generate the correlation finctions employed here~\cite{soviet} and/or coherence effects. 

Future efforts along these lines are
to be encouraged, as are complete calculations of the $n$-$n$ relative momenta based on realistic 
halo wavefunctions coupled to a reaction model and incorporating the effects of the FSI in the outgoing channel.  Ultimately, the comparison of such predictions for different wavefunctions with
the measured relative momentum distribution should provide the most reliable insight into the
halo neutron spatial configurations.

\section*{Acknowledgments}

I would like to underline the key r\^oles played by Miguel Marqu\'es (LPC-Caen) and 
Guillaume Normand (LPC-Caen)
in the work presented here.  Appreciation is also due to my colleagues in the long standing
LPC-CHARISSA-DEMON collaboration -- in particular, Martin Freer (Birmingham), Wilton Catford (Surrey), Louise Stuttg\'e (IPHC-Strasbourg) and Francis Hanappe (ULB-Bruxelles).  The support provided by the technical staffs of LPC-Caen and GANIL, together
with that of Alain Ninane (Slashdev Integrated Systems Solutions),
has been indespensable to the undertaking and running of the experiments.


\newpage

\end{document}